\documentclass[twocolumn,aps,prl]{revtex4}
\usepackage{amsmath}
\usepackage{graphicx}
\usepackage{esint}
\usepackage{epsfig}

\def\beq{\begin{equation}}
\def\eeq{\end{equation}}
\def\beqa{\begin{eqnarray}}
\def\eeqa{\end{eqnarray}}
\def\e{\epsilon}

\def\e{\epsilon}
\def\cH{{\mathcal H}}

\def\cG{{\mathcal G}}

\def\Tr{\mathrm{Tr}}
\def\r{{\bf r}}

\def\o{\omega}

\def\nonum{\nonumber \\}

\def\SAM{ \mathrm{SAM}}
\def\elec{ \mathrm{elec}}
\def\nonum{ \nonumber \\}
\renewcommand{\sec}[1]{\vskip 0.5truecm \noindent \emph{#1}. -- }

\begin{document}  
\author{Yonatan Dubi}
\email{jdubi@bgu.ac.il}
\affiliation{Department of Chemistry and the Ilze-Katz Institute for Nano-Scale Science and Technology, Ben-Gurion University of the Negev, Beer-Sheva 84105, Israel}
\title{Transport Through Self-Assembled Monolayer Molecular Junctions: Role of In-Plane Dephasing}
\date{\today}
\begin{abstract}
Self-assembled-monolayer (SAM) molecular junctions (MJs) constitute a promising building block candidate for future molecular electronic 
devices. Transport properties of SAM-MJs are usually calculate using either the phenomenological Simmons model, 
or a fully-coherent transport theory, employing the SAMs periodicity. We suggest that dephasing plays an important role in determining the transport properties
of SAM-MJs. We present an approach for calculating the transport properties of SAM-MJs that inherently takes into account in-plane dephasing in the electron motion as it 
traverses the SAM plane. The calculation is based on the non-equilibrium Green's function formalism, with a local dynamics approximation that describes incoherent 
motion along the SAM plane. Our approach describes well the two hallmarks of transport through SAM-MJs, namely the exponential decay of current with molecular 
chain length and the reduction of the current per molecule as compared to single-molecule junctions. Specifically, we show that dephasing leads to an exponential decay 
of the current as a function of molecular length, even for resonant tunneling, where the fully coherent calculation shows little or no length-dependence of the current. 
The dephasing is also shown to lead to a substantial reduction of the current in a SAM-MJ as compared to the single molecule junction, in a realistic parameter 
regime, where the coherent calculation shows only a very small reduction of the current. Finally, we discuss the effect of dephasing on more subtle transport phenomena 
such as the conductance even-odd effect and negative differential resistance.                                       
\end{abstract}

\maketitle   


\section{Introduction} Self-assembled-monolayer (SAM) based molecular junctions (MJs) may become the basic building block for future molecular organic electronics \cite{Akkerman2006,Akkerman2008,Aviram1974,Tyagi2011,Wang2005,Love2005,Joachim2000,Grimm2013,BofBufon2010,Ringk2013,Velessiotis2011,Aswal2006,Marya2003,Vijayamohanan2001,Vondrak2000, Vuillaume2008,Jager2013,Halik2011,Heath2009,Walker2013,Scheer2010}. The most common theoretical tool to analyze SAM-MJs transport properties 
is the so-called Simmons model, which essentially treats the MJs as a square potential barrier, typically taking the difference between the Fermi level in the electrodes
and the molecular orbital energy as the barrier 
height \cite{Wang2005a,Akkerman2007,Vilan2007,Akkerman2008,Luo2011,Kronemeijer2011,Chen2013}. Evidently, such a phenomenological model cannot capture all the subtleties 
that accompany SAM molecular structure \cite{Vilan2007}. For instance, it does not explicitly take into account the inter-molecular interactions, experimentally proven 
to have a large effect on the transport properties of SAM-MJs (see e.g. \cite{Nerngchamnong2013,Selzer2005}). A more microscopic approach to study transport properties of SAM-MJs is based on  calculating the transmission function of the SAM-MJ using Green's functions
 \cite{Meir1992,Datta1997a,DiVentra2008}, treating the SAM as a perfect periodic lattice \cite{Magoga1997,Tomfohr2004,Prodan2009,Landau2009,Landau2008,Andrews2008,Egger2012,George2008,Heimel2010,Kim2005,Leijnse2013,Reuter2011,Chen2013} and assuming that the SAM generate coherent electronic bands \cite{Tomfohr2002,Tomfohr2004}. 

There are two prominent hallmarks of transport through SAM-MJs. The first is the exponential decay of current (or conductance at low biases) with the number of 
molecules in the molecular chains which form the SAM (see e.g. \cite{Cui2002,Wold2001,Wold2002,York2003,Salomon2003}). Typically, the conductance decays exponentially with the chain length and is given by \beq G=G_0 e^{-\beta n} \label{cond-decay} ~~,\eeq where $G_0$ describes  
the tunneling from the electrodes into the molecule, $\beta$ is the exponential decay coefficient and $n$ the number of atoms in the molecule. $\beta$ is commonly discussed within the Simmons model, where the width of the barrier replaces the molecular length. This 
is clearly an over-simplification of the molecular junction \cite{Vilan2007}, since, for instance, $\beta$ does not take into account the detailed structure of the 
molecule and certainly not the SAM inter-molecular tunneling. The Simmons model was extended to account for inter-molecular coupling \cite{ Slowinski1997,Yamamoto2002} assuming fully incoherent electron tranfer between neighboring molecules. Within the microscopic theory, the exponential decay of length is explained as off-resonant tunneling \cite{Scheer2010}, when the Fermi energy lies outside the coherent band formed by the SAM (effectively making it similar to the Simmons model). 

This theory has several drawbacks. First, the exponential decay typically persists even at large biases where the energies may be beyond the band edges. 
Furthermore, it seems that the exponential decay is a universal feature, which may occur even if the Fermi energy lies within the molecular bands. Assuming 
intra-molecular coupling $t$ and inter-molecular coupling 	$\gamma$, the width of the molecular band would be $\sim 2t+4\gamma$, which means that to lie outside the 
molecular band the Fermi level has to lie at a distance $t+2\gamma$ from the molecular level. An estimate of intra-molecular coupling of $t \sim 1.5-3$ eV and a similar 
value for $\gamma$ would mean that the Fermi level has to lie at least $\sim 3-6$eV away from the molecular orbitals. Put differently, the fact that in a {\sl single 
molecule} junction the transport is off-resonance does not guarantee that it is the case also for transport through a SAM-MJ of the same molecule, due to the 
inter-molecular coupling.

The second hallmark of transport through SAM-MJs is the reduction of the current per molecule, as compared to a single-molecule junction with the same molecule. 
Experimentally, it has been shown that the current of a SAM-MJ is dramatically smaller (by 
three orders of magnitude) than for a single-molecule junction with the same molecules, a reduction that has been attributed to the molecular environment of the SAM 
\cite{Salomon2003,Selzer2005}. This was addressed by Landau {\sl et al.} \cite{Landau2008}, who showed a reduction in the transmission function of a SAM as compared to 
the single-molecule junction. As discussed in details in the supplemental material, we have repeated the calculation (using a wide-band approximation), essentially 
reproducing the results of Ref.~\cite{Landau2008}. However, the calculation shows that in order to achieve a three orders of magnitude reduction in the current, an 
unphysical inter-molecular coupling that exceeds $6$ eV is needed, again pointing to an inconsistency between the theory and experiment. 

Overcoming the above apparent inconsistencies is the main goal of this work. To do so, we posit that a central missing ingredient in the coherent band description 
is the dephasing in the electron motion as it traverses the SAM plane. This dephasing may occur due to the interaction of electrons with the soft surface phonon modes 
\cite{Rosenbaum2004}, that are not present in single-molecule junctions (which only have optical modes). These soft modes can strongly affect the electron 
motion (electron-phonon interaction strength is inversely proportional to the phonon frequency), but the electron-phonon scattering can be almost elastic, resulting only in the loss of 
electron phases during its motion in the SAM plane. Dephasing due to electron-phonon interactions (along the molecular chain) in single-molecule junctions was recently 
shown to lead to an exponential decay of conductance with length for short molecular wires \cite{Pal2011}. Earlier studies of dephasing in molecular junctions  demonstrated that the length-dependence of the conductance can critically depend on dephasing \cite{Segal2000,Segal2002}.

\begin{figure}[h!]

\includegraphics[width=9.5truecm]{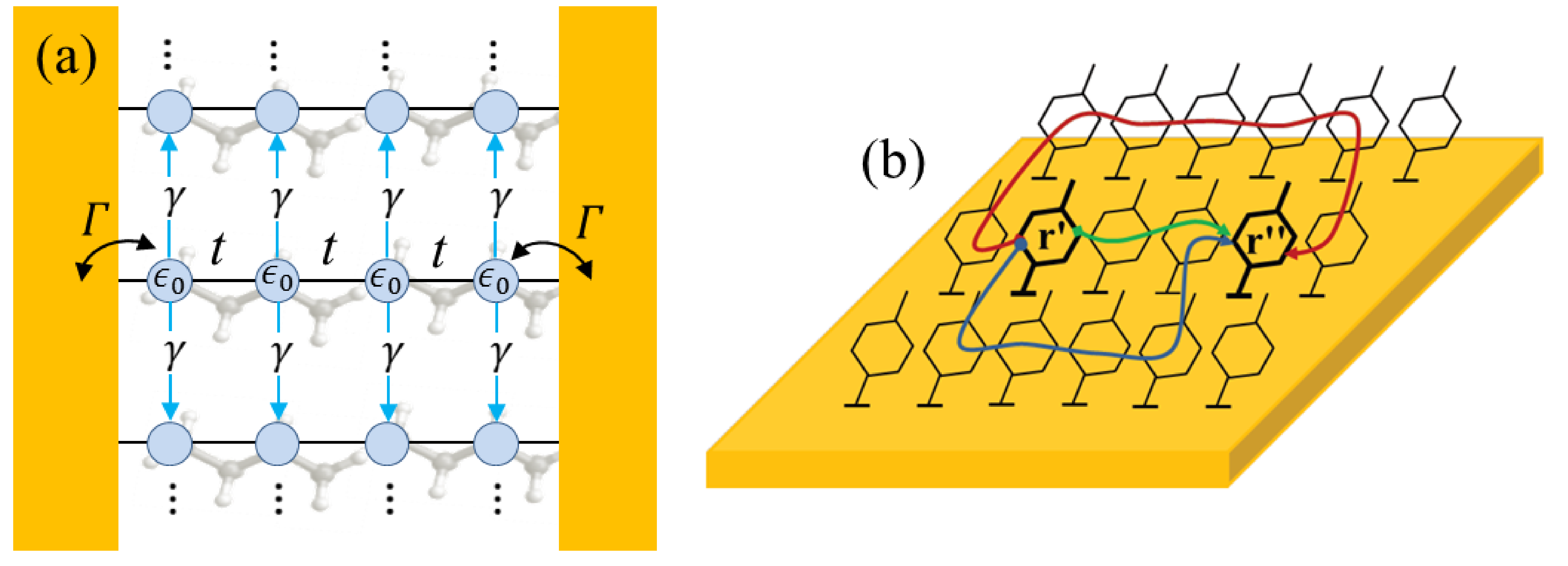}
\caption{(a) Schematic illustration of the model: A SAM of molecular chains with $n$ orbitals. The electrons in the SAM 
 can tunnel between the chain orbitals (with tunneling amplitude $t$), from the molecules into the electrodes (inducing a level broadening $\Gamma$) and between nearest-neighbor molecules in the SAM (with tunneling amplitude $\gamma$). This inter-molecular tunneling gives rise to a
 self-energy term $\Sigma_{\SAM}$ that is calculated self-consistently with the single-electron Green's function (Eq.~\ref{selfconst}). (b) Schematic presentation of the calculation leading to Eq.~\ref{selfconst}. The Green's function $\mathcal{G}_{\r',\r''; m,m'}$ in Eq.~\ref{Sigma1} describes 
the electron propagation from point $\r'$ to point $\r''$ in the SAM. The propagation can occur via different paths, three possible examples are shown (red, green and blue lines). The approximation now assumes that due to dephasing, each path carries a different random phase contribution, and these random phases average out to a vanishingly small contribution to the total Green's function.  }
\label{fig1}
\end{figure}

\sec{Model and method} We model the single molecule as a chain of orbitals 
connected by tunneling matrix elements (schematically depicted in Fig.~\ref{fig1}(a)), similar to the model used in, e.g., Refs.~\cite{Peskin2010,Landau2008,Landau2009}. The Hamiltonian for a molecular chain at position $\r$ in the SAM is 
given by 
\beq \cH_\r=\e_0 \sum^n_{i} |\r,i\rangle \langle \r,i |-t \sum^n_{i} |\r,i\rangle \langle \r,i+1 |+\mathrm{H.c.}~,
\eeq 
where $i$ is the position along 
the chain, $n$ is the number of orbitals(i.e. the chain length), $\e_0$ is the molecular orbital energy and $t$ is the intra-molecular hopping matrix element. The inter-molecular tunneling is 
assumed to takes place only between nearest-neighboring atoms, and only between orbitals at the same position $i$ along the molecule \cite{Landau2008,Landau2009}. The corresponding 
Hamiltonian term is
\beq
\cH_{\langle\r,\r'\rangle }=-\gamma \sum^n_{i} |\r,i\rangle \langle \r',i |+\mathrm{H.c.}~, \label{Ham}
\eeq 
where $\gamma$ is the inter-molecule hopping matrix element, which interpolates between the 
single-molecule junction ($\gamma=0$) and the SAM-MJ ($\gamma \neq 0$). 

The calculation is based on the Green's function (NEGF) theory \cite{DiVentra2008}, and the properties of the SAM as well as the dephasing are self-consistently incorporated into the molecular self-energy\cite{Anantram1995,Lake1992,Datta1990}. 
We treat electron dephasing on a phenomenological level, common in the study of dephasing effects on transport, see, e.g., Refs.~\cite{DAmato1990,Krems2009,Pal2011,Nozaki2012}. The method is presented here for a general Hamiltonian, and then implemented for the specific 
simplified model for the SAM of Eq.~(\ref{Ham}).

Within the Green's function approach, one evaluates the GF of the molecular junction 
\beq \mathcal{G}^{r,a}(\o) =\left( E - \mathcal{H}\pm \Sigma^{r,a} \right)^{-1} ~~, \label{GreenFunction1}\eeq  
where $\mathcal{G}^{r,a}(\o)$ are the retarded/advanced Green's function and $\Sigma^{r,a}$
 are the retarded/advanced self-energies 
due to the electrodes, which are assumed to be known (typically these are taken in the wide-band approximation \cite{DiVentra2008,Peskin2010}). Once  $\mathcal{G}^{r,a}(\o)$ is known, the current is given by \cite{Meir1992,Datta1997a,DiVentra2008}
\beq J=\frac{e}{h} \int \mathrm{d} E \Tr \left(\Sigma^r_L \mathcal{G}^r \Sigma^a_R \mathcal{G}^a \right) (f_L(E)-f_R(E)) \label{MeirWingreen} \eeq 
where $\Sigma_{L,R}$ is the contribution
 to the self-energy from the left/right electrode and $f_{L,R}$ are the Fermi functions of the left and right electrodes, respectively, 

$f_{L,R}(E)=\left(1+\exp(\frac{E-\mu_{L,R}}{T}) \right)^{-1}$. The chemical potentials are tuned by the external voltage bias, $\mu_{L,R}=\mu\pm V/2$ (hereafter we set $\mu=0$). The 
conductance of the junction (per molecule) is given by $G=\frac{e^2}{h}\tau(\mu)$, where $\tau(E)=\Tr \left(\Sigma^r_L \mathcal{G}^r \Sigma^a_R \mathcal{G}^a \right)$ is the transmission function.

For a microscopic description of transport through SAM-MJ, one has to consider a parallel arrangement of molecules that are in contact with the electrodes and with 
each other, and incorporate this into the GF.  To do so, we write the Hamiltonian as follows,
 \beq \cH_{\SAM}=\sum_{\r} \cH_\r +  \sum_{\langle \r,\r' \rangle } \cH_{\r,\r'} ~, \label{Hamiltonian1}
\eeq
where $\r,\r'$ are positions on the two-dimensional plane of the SAM, and we assume that only nearest-neighbor $\langle \r,\r' \rangle$ molecules interact. Writing the Hamiltonian in the basis of the single molecules 
we have 
$\cH_\r=\sum_n E_n c^{\dagger}_{\r,n} c_{\r,n}$, where $c^{\dagger}_{\r,n}$ is the creation operator for the molecular orbital $| \r,n \rangle $ in the molecule (with energy $E_n$) located at position $\r$. The inter-molecule tunneling Hamiltonian has the general form 
\beq 
\cH_{\r,\r'}=\sum_{n,m} \gamma^{n,m}_{\r,\r'} c^\dagger_{\r,n} c_{\r',m} + h.c~~. \label{Hamiltonian2}
\eeq 

To calculate the single-molecule Green's function $\mathcal{G}^{r}_{\r; n}$, we use the Dyson's equation, which for the Hamiltonian above gives
\beq
\mathcal{G}^{r}_{\r; n}=g^{r}_{\r;n}+g^{r}_{\r;n}\sum_{\r',m} \gamma^{n,m}_{\r,\r'} \mathcal{G}_{\r',\r; m,n} ~~,\eeq
where $g^{r}_{\r;n}$ is the bare Green's function (of the single molecule, no SAM). The off-diagonal full Green's function $\mathcal{G}_{\r',\r; m,n}$ 
can be expressed in terms of $\mathcal{G}^{r}$  with an additional Dyson's equation, leading to the term 
\beq 
\mathcal{G}_{\r',\r; m,n}=\sum_{\r'',m'} \gamma^{m',m}_{\r'',\r'} g_{\r',\r'; m,m}\mathcal{G}^{r}_{\r'',\r'; m',n}~. \eeq Plugging this back into the 
first Dyson's equation leads to the following expression, 
\beq 
\mathcal{G}^{r}_{\r}=\left( \omega - \cH_\r+\Sigma^r_{\elec}+\Sigma^r_{\SAM} \right)^{-1} \label{GreenFunction2}\eeq where the self-energy $\Sigma_\SAM$ takes into account the inter-molecule tunneling within the SAM, and $\Sigma_\elec$ is the self-energy due to the electrodes.
The self-energy (in matrix notation) is given by
\beq 
\Sigma_{\SAM} (\r; n)= \sum_{\r',\r'',m,m'} \gamma^{n,m}_{\r,\r'}\gamma^{m',n}_{\r'',\r} \mathcal{G}_{\r',\r''; m,m'}~~,\label{Sigma1}
\eeq 
where both $\r',\r''$ are nearest neighbors of $\r$ and the Green's function is defined as $\mathcal{G}_{\r',\r''; m,m'} \sim \langle c^\dagger_{\r',m} c_{\r'',m'} \rangle $. 
This description of the self-energy is not useful, since the non-local Green's function cannot be simply determined. 

To proceed, we posit that an electron  which tunnels from the molecule at $\r$ to its
 neighbor at $\r'$, will return to $\r$ from that same molecule at $\r'$. This is equivalent to the assumption that an electron traversing the SAM loses its phase when 
making more than one consecutive hop  between the molecules (or, put differently, to assuming that the dephasing time is shorter than the typical time it takes the 
electron to traverse between two neighboring molecules on the SAM). The Green's function $\mathcal{G}_{\r',\r''; m,m'}$ in Eq.~\ref{Sigma1} describes 
the electron propagation from point $\r'$ to point $\r''$ in the SAM. The propagation can occur via different paths ( three possible examples are shown in Fig.~\ref{fig1}(b) by red, green and blue lines). One can write then $\mathcal{G}_{\r',\r''; m,m'}=\sum_{p} |G_p(\r',\r'')| e^{i \phi_p(\r',\r'')} $, where the sum is over 
all the different paths $p$ leading from point $\r'$ to point $\r''$, $|G_p(\r',\r'')|$ is the amplitude of the given path and $\phi_p(\r',\r'')$ is the phase acquired 
by the electron wave function. Due to the dephasing, $\phi_p(\r',\r'')$  are random phases, and therefore their sum vanishes. 

The only case where the phases do not cancel out is the case where $
\r'=\r''$. Thus, that the main contribution to the SAM self-energy is from the local part, i.e.,
\beq \Sigma_{\SAM} (\r; n)=\sum_{\r', m} | \gamma^{n,m}_{\r,\r'}|^2 \mathcal{G}_{m,\r'}~~.\label{Sigma2}\eeq

Assuming that the SAM has spatial homogeneity, the Green's functions and the tunneling matrix elements do not depend on the positions, and one is left with 
\beqa
\mathcal{G}^{r}&=&\left( \omega - \cH_\r+\Sigma^r_{\elec}+\Sigma^r_{\SAM} \right)^{-1} \nonum
\left[ \Sigma^r_{\SAM}  \right]_{n,m}&=&| \gamma^{n,m} |^2 \mathcal{G}^r_{m,n}, ~~~ \mathcal{G}^r_{m,n}=\langle m| \mathcal{G}^r |n \rangle ~~, \label{selfconst} 
\eeqa
which are to be solved self-consistently (a Feynman diagram description of the derivation is given in the Supplemental material).



Now returning to the specific model discussed here, the specific choice for the  inter-molecular tunneling in the Hamiltonian of Eq.~\ref{Ham} leads to a SAM self-energy of the 
form $\left[ \Sigma^r_\SAM \right]_{i,j}=\delta_{i,j} \gamma^2 \langle i| \cG^r |i\rangle $.  The electrode self-energies 
$\Sigma^r_{\elec}=\Sigma^r_{\elec,L}+\Sigma^r_{\elec,R}$ (for the left and right electrodes, respectively), are taken in the wide band approximation \cite{Peskin2010}. 
Considering the entire SAM, the electrode self-energies would be 
\beqa
\Sigma^r_{\elec,L}=-i \frac{\Gamma}{2} \sum_{\r} |L (\r) \rangle\langle L (\r)|~, \nonum
\Sigma^r_{\elec,R}=-i \frac{\Gamma}{2} \sum_{\r} |R (\r) \rangle\langle R (\r)| ~,\label{Sigma}
\eeqa
where $\Gamma$ is the level 
broadening and $|L (\r)\rangle$ and $|R (\r)\rangle$ are the left-most and right-most orbitals of the molecular chain at position $\r$ along the SAM plane. However, 
for the single-molecule nature of our approach and assuming that the SAM and electrodes are homogeneous, the self-energy becomes independent of position 
and in terms of the single molecule orbital, it is given by 
$\Sigma^r_{\elec,L}=-i \frac{\Gamma}{2} |L\rangle\langle L|, \Sigma^r_{\elec,R}=-i \frac{\Gamma}{2} |R\rangle\langle R|~ $ (see Fig.~\ref{fig1}(a)).

\section{Results - length dependence of conductance and current reduction} We start by studying the length-dependence of the conductance. In Fig.~\ref{fig1}(b) the transmission as a function of molecule length $n$ is plotted for 
different values of the inter-molecule tunneling amplitude $\gamma=0,0.1,0.2,...,2$ eV (other parameters are $\e_0=0.3$ eV, $\Gamma=50$ meV and $t=1.6$ eV). It is important to note 
that we are not aiming at a detailed quantitative agreement with some specific experiments, but rather, to demonstrate qualitative agreement with known results. Thus, the parameters we chose 
should be understood to represent a general example rather than a specific system. A clear 
crossover is observed from a ballistic regime, characterized by an oscillating dependence of conductance on length, to an exponentially decaying transmission (note the 
log scale). We note that by "ballistic transport" we mean that once an electron 
jumps from the electrode to the molecule, it does not scatter until it reaches the other side of the junction. Note that the Fermi energy is well within the molecular 
band, and for a fully coherent model would result in an oscillatory dependence of length. We thus conclude that electron dephasing directly leads to an exponential decay of conductance, even in resonant transport (i.e. when the Fermi level is within the molecular band) \cite{Pal2011}. 

It is useful to describe our system in the terms presented in Refs.~\cite{Segal2000,Segal2002}, which demonstrated that dephasing (due to, e.g., molecule-solvent 
interactions) can change the length-dependence of the conductance, driving it from a coherent dependence ("tunneling regime") to a normal ohmic dependence ("hopping 
regime"), a crossover which was also shown to happen in thermal transport of one-dimensional systems \cite{Dubi2009}. The SAM-MJ in the presence of in-plane dephasing is 
thus a system which displays am in-plane "hopping" behavior coupled to a longitudinal "tunneling" behavior, resulting in an exponential dependence of the current on 
chain length even at the resonant tunneling regime.

\begin{figure}[h!]
\centering
\includegraphics[width=6.5truecm]{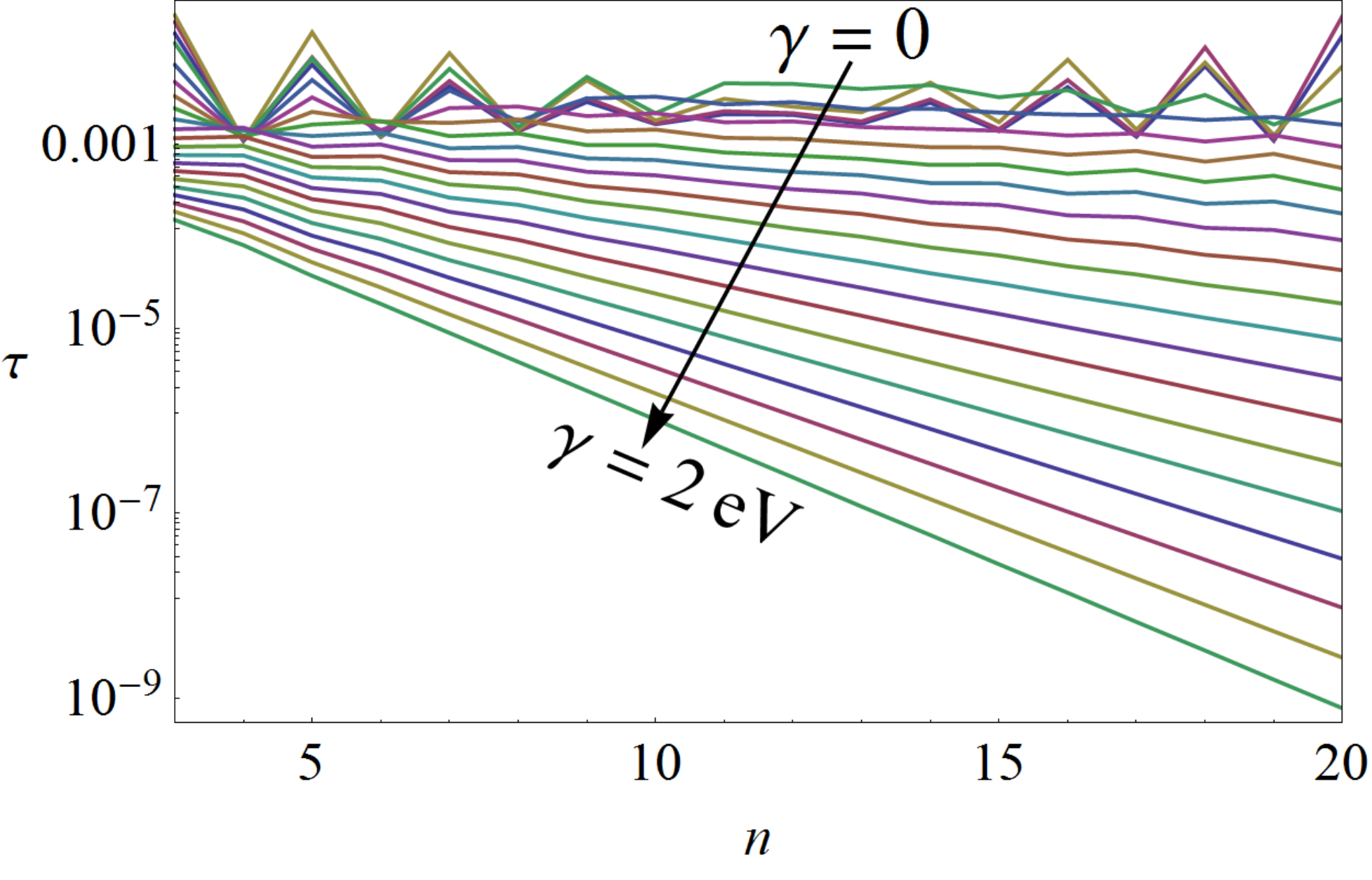}
\caption{Transmission as a function of molecule length $n$ for 
different values of the inter-molecule tunneling amplitude $\gamma=0,0.1,0.2,...,2$ eV, displaying a crossover from ballistic transport to exponential decay.}
\label{fig2}
\end{figure}

\begin{figure}[h!]
\centering
\includegraphics[width=8.0truecm]{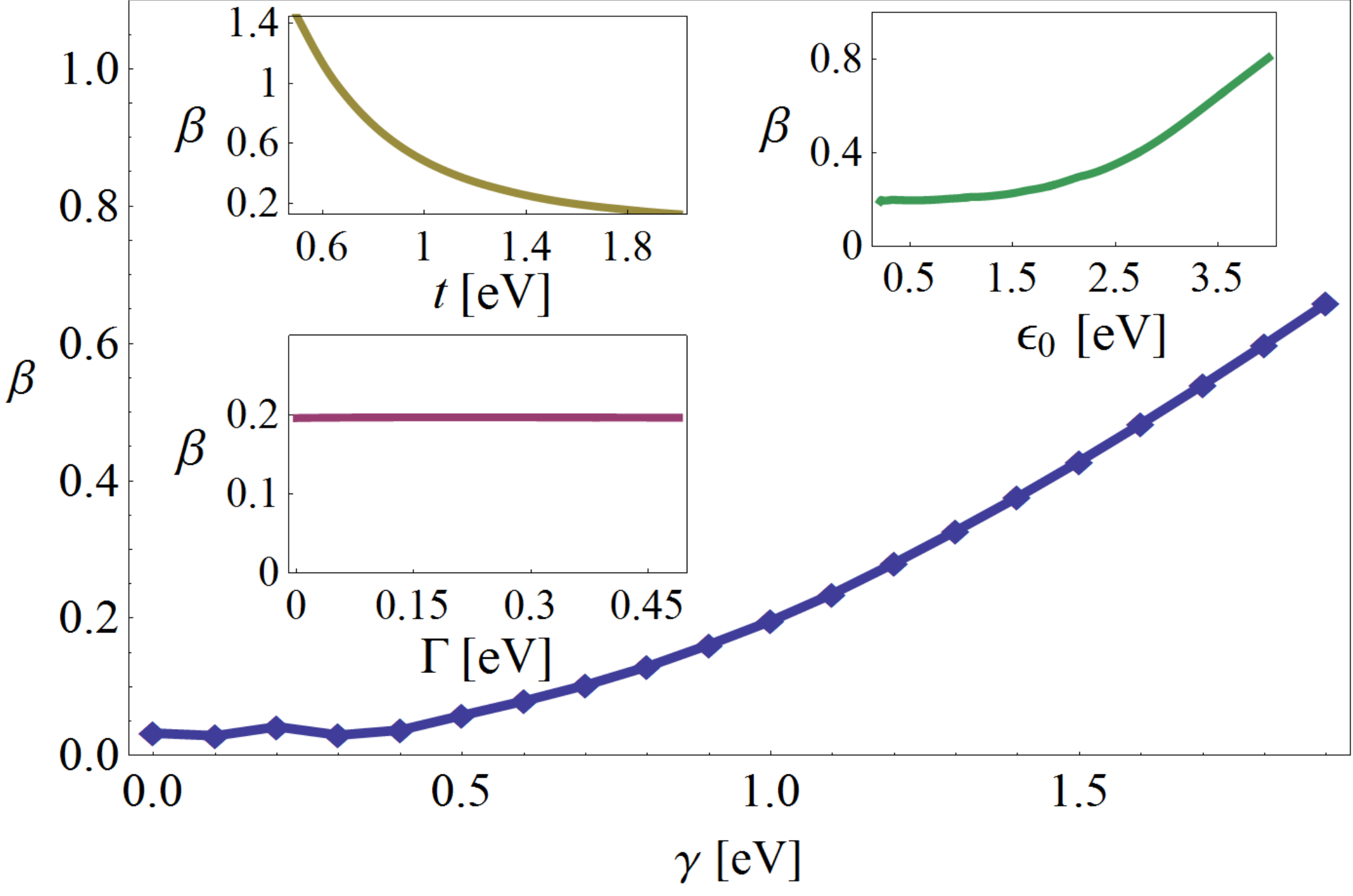}
\caption{Exponential decay coefficient $\beta$ as a function molecular junction parameters. Main panel: $\beta$ as a function of the inter-molecular tunneling amplitude $\gamma$. Top-left inset: $\beta$ Vs. intra-molecule tunneling amplitude $t$. Top-right inset: $\beta$ Vs. molecular orbital energy $\e_0$. Bottom inset: $\beta$ Vs. level broadening $\Gamma$. }
\label{fig3}
\end{figure}

The method allows us to directly relate the exponential decay coefficient $\beta$ and the conductance coefficient $G_0$ defined in Eq.~(\ref{cond-decay}) (per molecule, in units of $e^2/h$) to the parameters of the molecule and the interface. 
In Fig.~\ref{fig3}, $\beta$ is plotted as a function of the various parameters describing the molecular junction, namely, the inter-molecule coupling $\gamma$ (main panel), the molecular orbital energy level 
$\e_0$ (top-right inset), the intra-molecular tunneling amplitude $t$ (top-left inset) and the electrode-induced level broadening $\Gamma$ (bottom inset). For each calculation, only one parameter 
is changed and the rest are kept fixed, with the numerical values $t=1.6$ eV, $\e_0=0.3$ eV, $\gamma=1$ eV and $\Gamma=0.05$ eV. Even this simple example shows several important features: (i) The 
obtained values of $\beta\sim 0.5-1.5$ range within the observed experimental values, (ii) $\beta$ strongly depends on the inter-molecule coupling $\gamma$, a dependence which 
has not been studied in detail in the literature, (iii) the dependence on $\e_0$ is strong, and there 
is almost no dependence on $\Gamma$, in accordance with the Simmons model.  The results of Fig.~\ref{fig2}-\ref{fig3} should be contrasted with the fully coherent 
transport calculation (as in Ref. \cite{Landau2008}) which does not take into account any dephasing, and for similar parameters cannot capture the exponential decay at all , as is explicitly shown in Fig. S1(b) of the supplemental material. 

\begin{figure}[h!]
\centering
\includegraphics[width=8truecm]{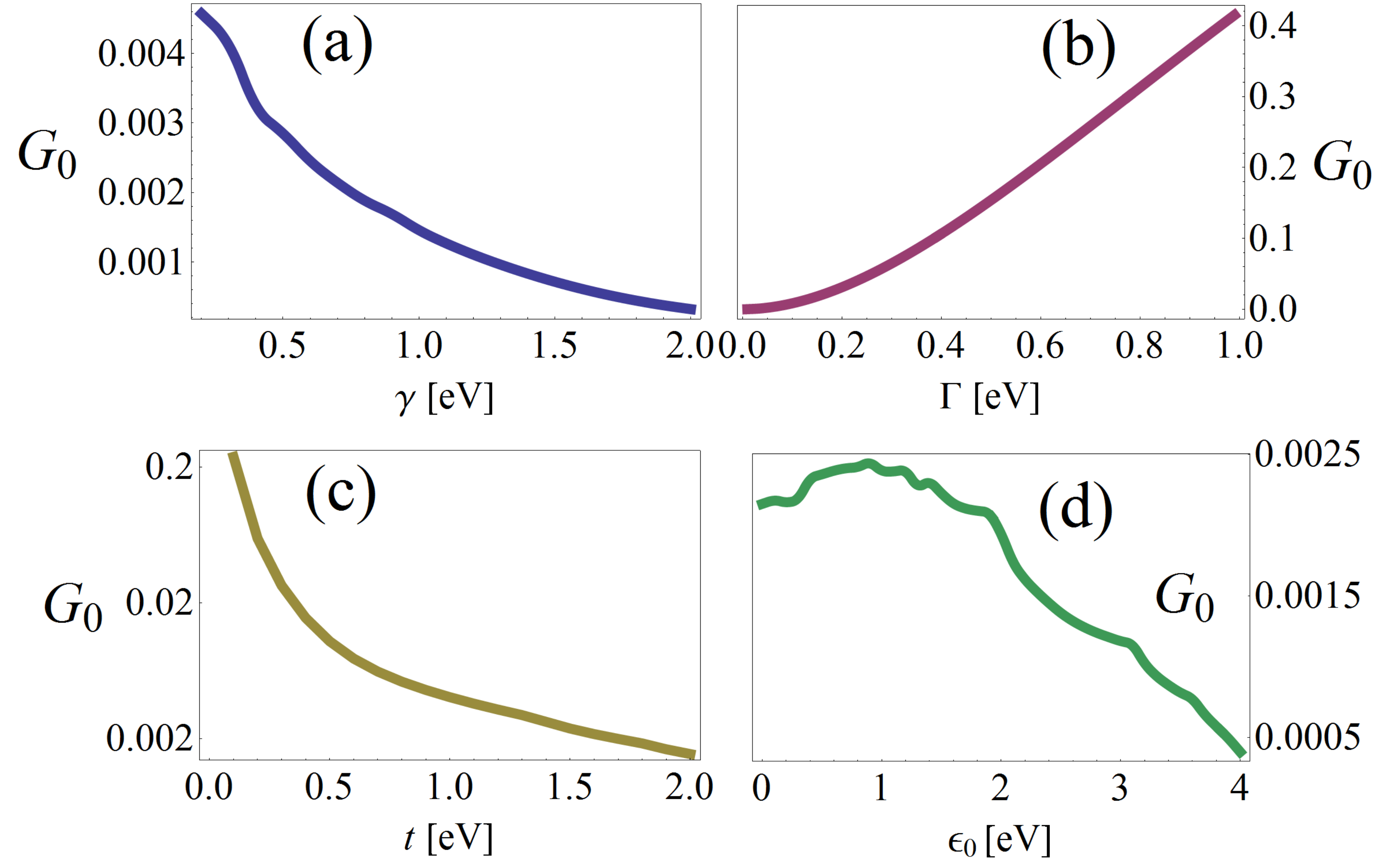}
\caption{$G_0$ (defined in Eq.~(\ref{cond-decay})) as a function of (a) inter-molecular coupling $\gamma$, (b) level broadening $\Gamma$, (c) intra-molecular hopping $t$ (log scale) and (d) molecular orbital energy $\epsilon_0$. }
\label{fig4}
\end{figure}

The conductance coefficient $G_0$ (per molecule, in units of $e^2/h$), can vary substantially between different SAM-MJs \cite{Cui2002,Wold2001,Wold2002,York2003}. In Fig.~\ref{fig4}, $G_0$ is plotted as a function of (a) inter-molecular coupling $\gamma$, (b) level broadening $\Gamma$, (c) intra-molecular hopping $t$ (in log scale) 
and (d) molecular orbital energy $\epsilon_0$ (numerical parameters same as in Fig.~\ref{fig3}). Notably, $G_0$ decays exponentially with $\gamma$, an effect that has never been 
discussed in the literature, and demonstrates the strong influence of inter-molecular coupling and dephasing on the transport properties of SAM-MJs. In addition, 
the strong dependence of $G_0$ on the other molecular parameters may explain the variations of $G_0$ between different SAM-MJs.

\begin{figure}[h!]
\centering
\includegraphics[width=7.0truecm]{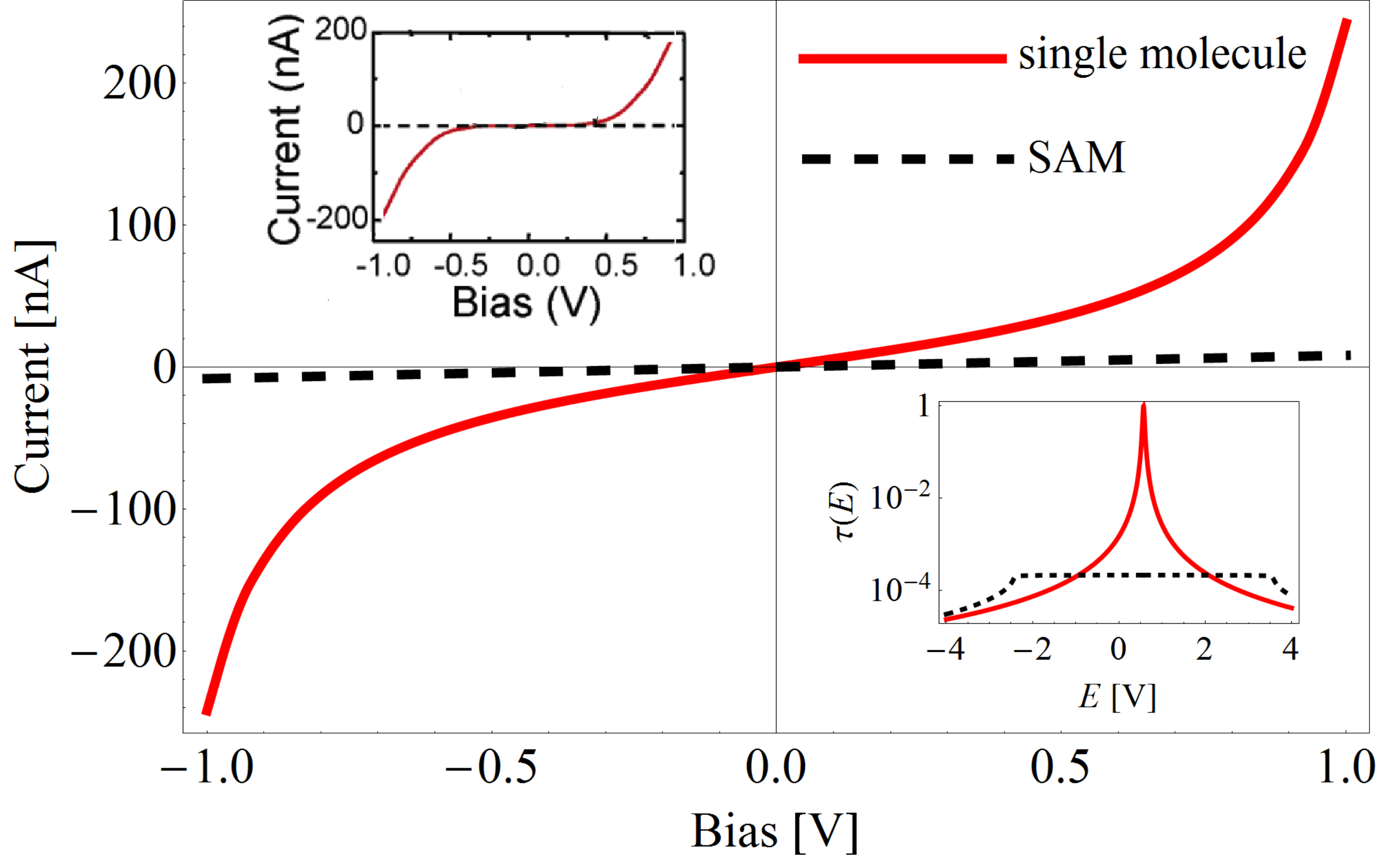}
\caption{Comparison between the I-V characteristics (per molecule) of a single-molecule junction (solid red line) and SAM-MJ (dashed black line), showing a huge decrease in the current 
though the SAM-MJ. Top-left inset: experimental results of the same, taken with permission from \cite{Selzer2005}. Bottom-right inset: transmission function $\tau(E)$ for the single-molecule and 
SAM-MJ , showing the origin of the decrease in current. See text for numerical parameters.}
\label{fig5}
\end{figure}

We next turn to the reduction of of current in SAM-MJs compared to their single-molecule counterparts. For the theoretical model, we consider a SAM where each molecule 
is described by a single orbital (i.e., $n=1$) \cite{Landau2008}. For this case, there is only one orbital and the  molecular Green's function 
obeys the self-consistency equation $\cG^r=\frac{1}{\o -\e_0-i \Gamma+4 \gamma^2 \cG^r}$ 
(we assume that the SAM is arranged in a square lattice with 4 nearest neighbors). From the Green's function one can calculate the transmission function and the current. In Fig.~\ref{fig5} the I-V characteristics of the junctions are shown 
for the two cases of (i) a single-molecule junction, i.e., $\gamma=0$, with no inter-molecule tunneling (solid red line), and (ii) a SAM-MJ with $\gamma=0.75$ eV (dashed black line). Other 
numerical parameters are the level 
broadening $\Gamma=20$ meV, molecular orbital energy $\e_0=0.5$ eV, and temperature $T=10$ K (which is experimentally relevant \cite{Selzer2005}, taken hereafter). These parameters were chosen to obtain a qualitative fit with the experiments. The experimental I-V measurements 
\cite{Selzer2005} are shown on the top-left inset for comparison, and the resemblance between theory and experiment is evident even for this extremely simple model for the molecule.

 To understand the origin of this huge reduction in the current for a SAM-MJ compared to the single-molecule junction, in the bottom-right inset we plot the transmission function $\tau(E)$ for the single-molecule and the SAM. For the single molecule junction, the transmission function has the usual Lorentzian shape and reaches $\tau=1$ at the resonance. For the SAM, on the other hand, the transmission exhibits a plateau with a low transmission value, about three orders of magnitude lower than the single molecule. This, again, should be contrasted 
with the fully coherent model, which requires an unrealistically large coupling $\gamma > 6$eV in order to reach the current reduction observed experimentally (see Fig. S3 in the supplemental material). We note that the reduction of current in SAM-MJs occurs assuming that the other 
molecular parameters (e.g., energy levels, coupling to electrode) remain the same.

This "flattening" of the transmission 
function, which also occurs for longer molecules, may help to explain the apparent robustness of SAM-MJs 
against fluctuations in the molecular parameters (mainly the positions of the HOMO and LUMO levels)\cite{Akkerman2006,Song2007,Thuo2011}. In single-molecule 
junctions, the conductance (and even the thermopower) are subject to large fluctuations \cite{Basch2005,Jang2006,Lortscher2007,Malen2009,Dubi2011,Dubi2013}, due to 
strong sensitivity of the transmission function to variations in both the molecular orbital energy and the molecule-electrode coupling (represented here by $\e_0$ and $\Gamma$, 
respectively). The HOMO energy can vary on a scale of $\sim 1$ eV due to slight changes in molecular alignment with the electrode. SAM-MJs, on the other hand, seem to 
be much more stable against these fluctuations, and the current distributions are significantly narrower than for single-molecule junctions. The flattened transmission 
function of SAM-MJs may offer a possible explanation, since changes in the position of the LUMO do not affect much the transmission (see supplemental material). 

The flattening of the transmission function has an additional consequence, as it provides the prediction that the thermopower of 
SAM-MJs would be substantially smaller than the thermopower of single-molecule junctions. the reason is that, at least at low temperatures, the thermopower $S$ is 
proportional to the logarithmic derivative of the transmission \cite{Dubi2011}, $\left. S\propto \frac{d \log \tau(E)}{dE} \right|_{E=E_f}$ (where $E_f$ is the Fermi 
energy). Thus, the flattening of the transmission function will result in a reduction of $S$, which may even change sign between the single-molecule and the SAM 
junctions. A detailed analysis of thermopower of SAMs is left for a future publication. 

\section{Results - Odd-even effect and negative differential resistance }
The dephasing mechanism we present here can also account for the more subtle features of transport through SAM-MJs. An example is the odd-even effect, which was recently demonstrated experimentally \cite{Thuo2011,Nerngchamnong2013}: the 
total current at a given voltage (or equivalently the conductance) shows the usual exponential decay with respect to the number or molecular elements (number of 
ethylenes in Ref. \cite{Thuo2011}), $j=j_0 e^{-\beta n}$, yet there is a statistically clear distinction between chains with odd or even number of ethylenes. Thuo {\sl et al.}~\cite{Thuo2011} do not support any specific origin for these oscillations, only stating that it is not captured by the Simmons model, thus pointing at its incompleteness. 

\begin{figure}[h!]
\centering
\includegraphics[width=7.0truecm]{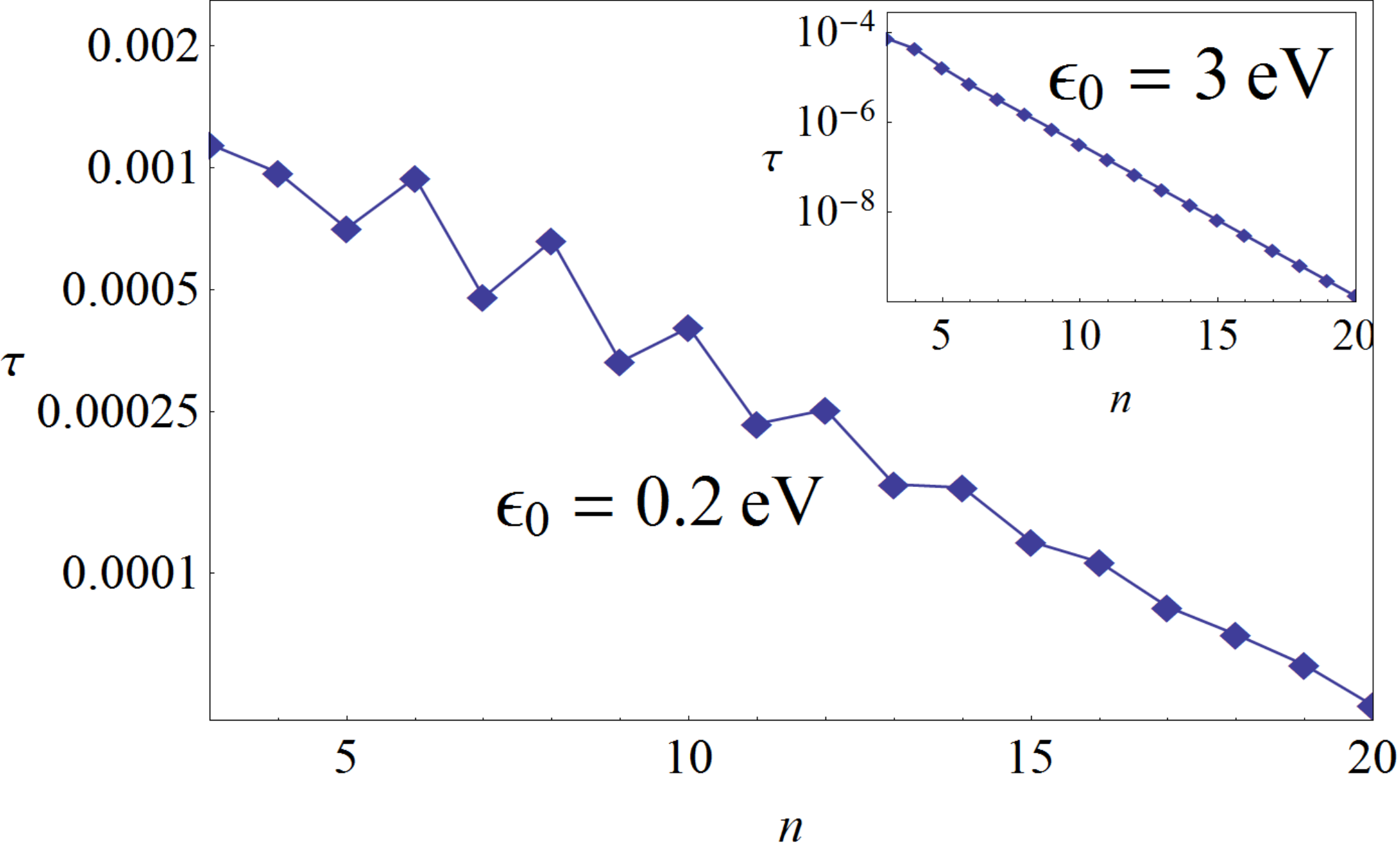}
\caption{Transmission $\tau$ as a function of molecular length $n$, for $\gamma=0.5 $eV and $\e_0=0.2$ eV. A clear odd-even effect is visible. Inset: same for $\e_0=3$ eV, indicating 
that the odd-even effect vanishes for molecular orbital  energy values that are far from the chemical potential.}
\label{fig6}
\end{figure}

In transport through ballistic molecular wires, one expects the transmission to have an odd-even effect due to the standing-wave nature of the electronic states in the wire \cite{Nozaki2008}. It turns out that this signature of coherent 
transport can persist even as $\gamma$ extrapolates from the molecular-wire ($\gamma=0$) to the SAM (large $\gamma$), depending on other parameters of the molecular junction. In Fig.~\ref{fig6} the transmission as a function of molecular length $n$ is plotted for $\gamma=0.5$eV. This is
a large value, considering that the level broadening is taken to be $\Gamma=0.05$ eV and the intra-molecular tunneling amplitude $t=1.6$eV. The odd-even effect is clearly visible, on top of the exponential decay of the transmission (note the log scale). Here the molecular orbital level is taken to be relatively close to the chemical potential (resonant tunneling), 
$\e_0=0.2$ eV. When considering a molecular orbital level that is further away from the chemical potential, however, the odd-even effect vanishes, as is shown in the inset of Fig. ~\ref{fig6}, where the transmission as a function of molecular length is plotted for $\e_0=3$ eV, and the odd-even 
effect cannot be detected. The odd-even effect is thus due to the delicate competition between coherent transport in the direction of the molecules and dephasing along the SAM plane. It is important to note that the explanation proposed here may not be the only cause for an odd-even effect: differences in the end-groups, contact angle and contact area also play roles in the odd-even effect. \cite{Nerngchamnong2013}

As a final example for the effect of dephasing on transport in SAM-MJs, we consider the negative differential resistance (NDR). When a 
molecular junction is placed under a voltage bias, the current is typically a monotonically increasing function of bias (resulting in the usual positive differential 
resistance). However, in certain situations, the current decreases with increasing bias, resulting in the so-called NDR. This effect has been observed in a large variety of 
molecular junctions, and it can be chemical in origin (i.e. due to chemical changes in the junction at some voltage) or electronic in origin \cite{Chen1999,Chen2000,Salomon2004,Selzer2002,Cheraghchi2008,Dalgleish2006,Kaasbjerg2011,Kang2010,Le2003,Leijnse2011,Long2008,Pati2008,Xu2008,Zhou2013,Zimbovskaya2008,Dubi2013a}. 

\begin{figure}[h!]
\centering
\includegraphics[width=8.5truecm]{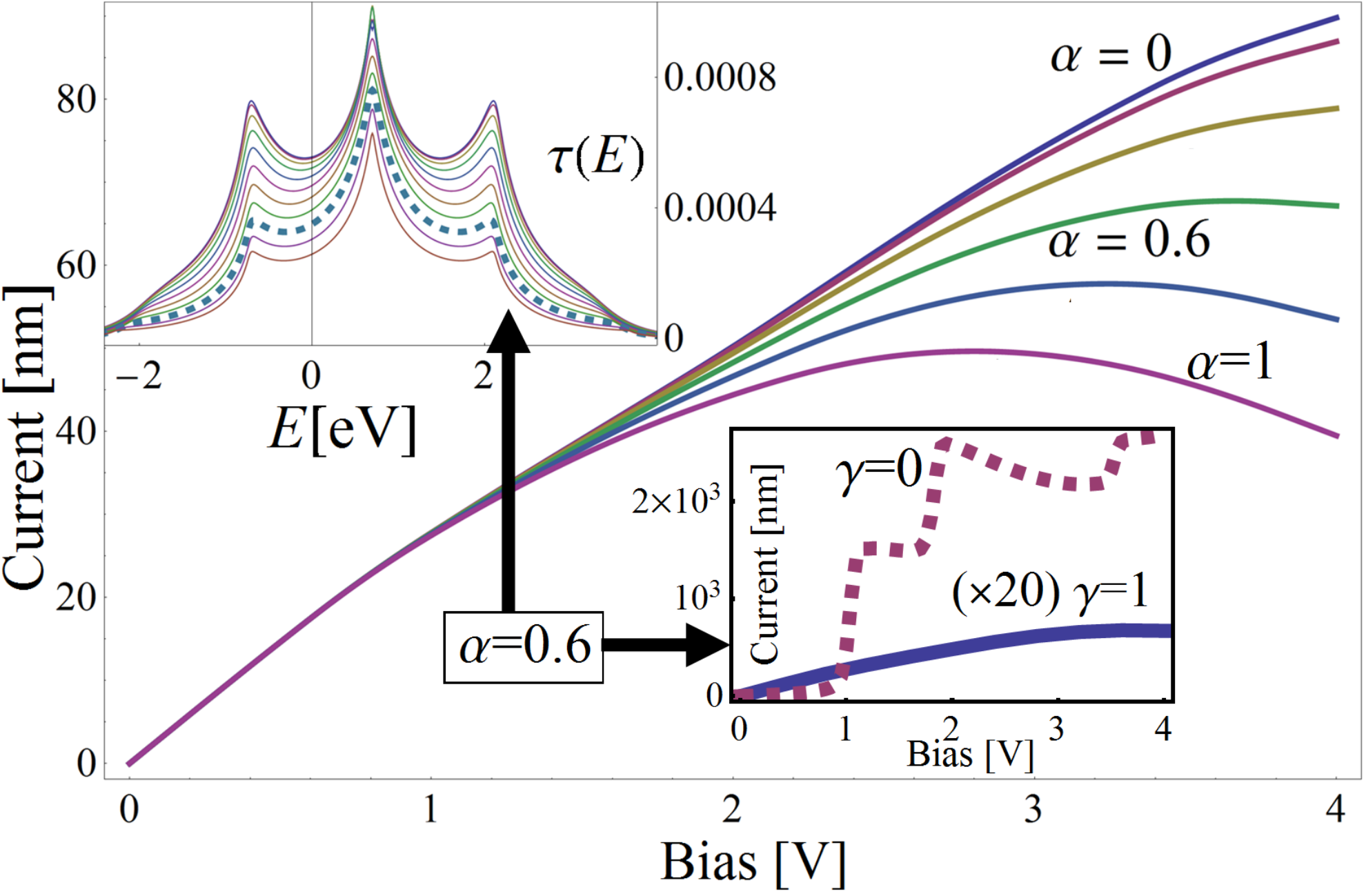}
\caption{I-V characteristics of a molecular chain of length $n$, for different values of the voltage drop along the molecule, $\alpha=0,0.2,0.4,...,1$. Top-left inset: transmission function for different values of voltage bias. Bottom-right inset: I-V characteristics for of a single-molecule junction ($\gamma=0$) and SAM-MJ ($\gamma=1$), for $\alpha=0.6$. }
\label{fig7}
\end{figure}

To demonstrate that our method and model can capture NDR, we add an additional term to the simple model of Eq.~\ref{Ham} to 
account for the voltage drop along the junction, $\cH_V=\sum^n_{i=1} \alpha V 
\left(\frac{1}{2}-\frac{i-1}{n-1} \right)|i\rangle \langle i |  $,~ so that there is a linear voltage drop of $\alpha V$ along the junction. The parameter $\alpha$ determines the fraction of voltage that 
drops on the molecule. For $\alpha=0$ all the voltage drop occurs at the contact point between the molecules and the electrodes, and for $\alpha=1$ the 
voltage drop is fully along the molecules. In principle, the voltage drop should 
be calculated self-consistently \cite{Mujica2000,Mujica2001,Ke2004,Zhang2005,Toher2008,Xue2003}. However, for the sake of simplicity and since we only wish to demonstrate the method's capabilities, we take $\alpha$ as an external parameter. 
In Fig.~\ref{fig7} the I-V characteristics of a molecular wire of length $n=6$ are plotted for different values of $\alpha=0,0.2,0.4,...,1$, ranging from no voltage drop on the molecule ($\alpha=0$) to a full potential drop on the molecule and no voltage drop along the molecule-electrode interface ($\alpha=1$). For large enough $\alpha$ a reduction 
of the current with increasing voltage, i.e. NDR, is observed.

 To track the origin of the NDR, in the top left inset the transmission function is plotted for different values of 
voltage $V=0,0.4,0.8,...,4$ eV (for $\alpha=0.6$), and the decrease in the total transmission is seen. The origin of the NDR is thus the following: the current is defined as the integral of the transmission function over 
the voltage drop window. As the voltage bias increases, the integration window increases, but simultaneously the transmission function decreases, and these are two competing 
effects. For low voltages the increase in current is due to the increase in the integration window. At a certain voltage which depends on the molecular parameters (marked in dashed 
line in the top-left inset), the two effects balance each other, leading to a maximum in the current. Beyond this bias, the reduction of the transmission due to the voltage drop 
along the junction becomes dominant, resulting in a decrease in the current and NDR. 

The NDR appears in single-moledcule junctions as well as in SAM-MJ. In the bottom-right inset of Fig.~\ref{fig7} we plot the I-V 
characteristics for molecular wire with $\gamma=0$ (single-molecule) and $\gamma=1$eV (SAM, current multiplied by a factor of twenty for comparison), for $\alpha=0.6$. 
The clear difference between the curves demonstrates the effect of the in-plane dephasing on the NDR, and the difference between transport though a single-molecule 
junction and SAM-MJ is a prediction which is verifiable within current experimental capabilities .  

\section{Conclusion}
In summary, we have shown that incorporating in-plane dephasing into the transport calculation allows us to explain several universal transport features of SAM-MJs. The current 
per molecule was calculated within the Green's function approach, and dephasing was treated on a phenomenological level, by self-consistently evaluating the contribution of the SAM to the self-energy. This method allowed us to address the exponential decay of current with molecular chain length, which was found to persist 
even when the Fermi energy lies within the molecular band, in contrast to the fully coherent approach (i.e. without dephasing). The in-plane dephasing was shown to lead 
to a substantial reduction of the current from the single-molecule to the SAM, which in the fully coherent approach would require unrealistically large inter-molecular 
coupling. Finally, the effect of dephasing on more subtle features such as the odd-even effect and negative differential resistance was considered.

We conclude by pointing out that the method introduced here, which inherently incorporates dephasing, can be simply implemented within quantum chemistry methods (and density-functional theory (DFT) in particular) which have been 
employed to study transport through single-molecule junctions. Since DFT methods require self-consistent calculation, our method does not add much to the overall 
computational cost. The intra-molecular coupling could be deduced from the orbital structure of the single-molecules, or from auxiliary calculations of a few 
molecules. This method is thus a way to introduce dephasing into {\sl ab initio} calculations, and paves the way for the first-principle study, prediction and 
design of the transport, thermoelectric \cite{Malen2010,Dubi2011}, optical \cite{Galperin2012} but also other \cite{Aradhya2013} properties of SAM-based molecular 
junctions that can be studied for single-molecule junctions.     

\acknowledgements
The author is grateful to I. Visoli-Fisher and C. Nijhuis for valuable discussions and comments. This research was funded by a BGU startup grant.
\providecommand*\mcitethebibliography{\thebibliography}
\csname @ifundefined\endcsname{endmcitethebibliography}
  {\let\endmcitethebibliography\endthebibliography}{}

\end{document}